\documentclass[pre,aps,twocolumn]{revtex4}
\usepackage{graphicx,epsf,subfigure,amsbsy,amsmath,float}
\usepackage[latin1]{inputenc}

\newcommand{\ket}[1]{| #1 \rangle}

\begin{document}

\title{Evaluating the locality of intrinsic precession damping in transition metals}

\author{Keith Gilmore$^{1,2}$ and Mark D. Stiles$^1$}

\affiliation{$^1$ Center for Nanoscale Science and Technology \\
National Institute of Standards and Technology, Gaithersburg, MD 20899-6202 \\
$^2$ Maryland Nanocenter, University of Maryland, College Park, MD 20742-3511 \\ }

\date{\today.}

\begin{abstract}

The Landau-Lifshitz-Gilbert damping parameter is typically assumed to be a local quantity, 
independent of magnetic configuration.  To test the validity of this assumption we 
calculate the precession damping rate of small amplitude non-uniform mode magnons in iron, cobalt, 
and nickel.  At scattering rates expected near and above room temperature, little 
change in the damping rate is found as the magnon wavelength is decreased from 
infinity to a length shorter than features probed in recent experiments.  This result indicates 
that non-local effects due to the 
presence of weakly non-uniform modes, expected in real devices, should not appreciably 
affect the dynamic response of the element at typical operating temperatures.  Conversely, at
scattering rates expected in very pure samples around cryogenic temperatures, non-local effects 
result in an order of magnitude 
decrease in damping rates for magnons with wavelengths commensurate with domain wall 
widths.  
While this low temperature result is likely of little 
practical importance, it provides an experimentally testable prediction 
of the non-local contribution of the spin-orbit torque-correlation model of 
precession damping.  None of these results exhibit strong dependence on the 
magnon propagation direction.  

\end{abstract}


\maketitle


Magnetization dynamics continues to be a technologically important, but
incompletely understood topic.  Historically, field induced magnetization 
dynamics have been described adequately by the phenomenological 
Landau-Lifshitz (LL) equation \cite{Landau.Lifshitz:1935}
~
\begin{equation}
\dot{\bf m} = -|\gamma_{\rm M}| {\bf m} \times {\bf H} + \lambda \hat{\bf m} \times \left (
{\bf m} \times {\bf H} \right ) \, ,
\label{LLG}
\end{equation}

\noindent or the mathematically equivalent Gilbert form \cite{Gilbert:1956, Gilbert:2004}.  Equation 
\ref{LLG} accounts for the near equilibrium dynamics of systems in the
absence of an electrical current.  $\gamma_M$ is the gyromagnetic ratio and $\lambda$
is the phenomenological damping parameter, which quantifies the decay of the
excited system back to equilibrium.  The LL equation is a rather simple approximation
to very intricate dynamic processes.  The limitations of the approximations entering into the LL 
equation are likely to be tested by the next generation of magnetodynamic devices.  While many 
generalizations for the LL equation are possible, we focus on investigating the importance of 
non-local contributions to damping.  It is generally assumed in both analyzing experimental results and in 
performing 
micromagnetic simulations that damping is a local phenomenon.  While no clear experimental evidence 
exists to contradict this assumption, the possibility that the damping 
is non-local -- that it depends, for example, on the local gradient of the magnetization -- would have 
particular implications for experiments that quantify spin-current polarization \cite{Vlaminck:2008}, 
for storage \cite{Parkin.racetrack:2008} and logic \cite{Parkin.shiftregister:2008} devices based on 
using this spin-current to move domain-walls, quantifying vortex \cite{Slonczewski:1984} and mode 
\cite{McMichael:2008} dynamics in patterned samples, and the behavior of nano-contact 
oscillators \cite{Rippard:2005, Mancoff:2005}.

While several viable mechanisms have been proposed to explain the damping process in different 
systems \cite{Kambersky:1967, Kambersky:1970, Korenman.Prange:1972,
Kambersky:1976, Tserkovnyak:2002, MacDonald:2004, Kelly:2005},
we restrict the scope of this paper to investigating the degree to which the assumption of local damping
is violated for small amplitude dynamics within pure bulk transition metal systems where the dominant 
source of damping is the intrinsic spin-orbit 
interaction.  For such systems, Kambersk\'{y}'s \cite{Kambersky:1976} spin-orbit torque-correlation 
model, which 
predicts a decay rate for the uniform precession mode of
~
\begin{equation}
\lambda_0 = \frac{\pi \hbar \gamma_{\rm M}^2}{\mu_0} \sum_{nm} \int d{\bf k} \left | \Gamma^-_{nm}({\bf 
k}) \right |^2 W_{nm}({\bf k}) \, ,
\label{Kambersky.original}
\end{equation}

\noindent has recently been demonstrated to account for the majority of damping 
\cite{Gilmore:2007, Kambersky:2007}.  The matrix elements $|\Gamma^-_{nm}({\bf k})|^2$ 
represent a scattering
event in which a quantum of the uniform mode decays into a
single quasi-particle electron-hole excitation.  This annihilation of a magnon
raises the angular momentum of the system, orienting the magnetization closer to
equilibrium.  The excited electron, which has wavevector ${\bf k}$ and band index
$m$, and the hole, with wavevector ${\bf k}$ and band index $n$, carry off the
energy and angular momentum of the magnon.  This electron-hole pair is rapidly
quenched through lattice scattering.  The weighting function $W_{nm}({\bf k})$
measures the rate at which the scattering event occurs.  The very short lifetime
of the electron-hole pair quasiparticle (on the order of fs at room temperature)
introduces significant energy broadening (several hundred meV).  The weighting 
function, which is a generalization of the delta function appearing in a simple Fermi's golden rule 
expression, quantifies the energy overlap of the broadened electron and hole states with each other and 
with the Fermi level.

\begin{figure}[t]
 \includegraphics[angle=0,width=8.0cm]{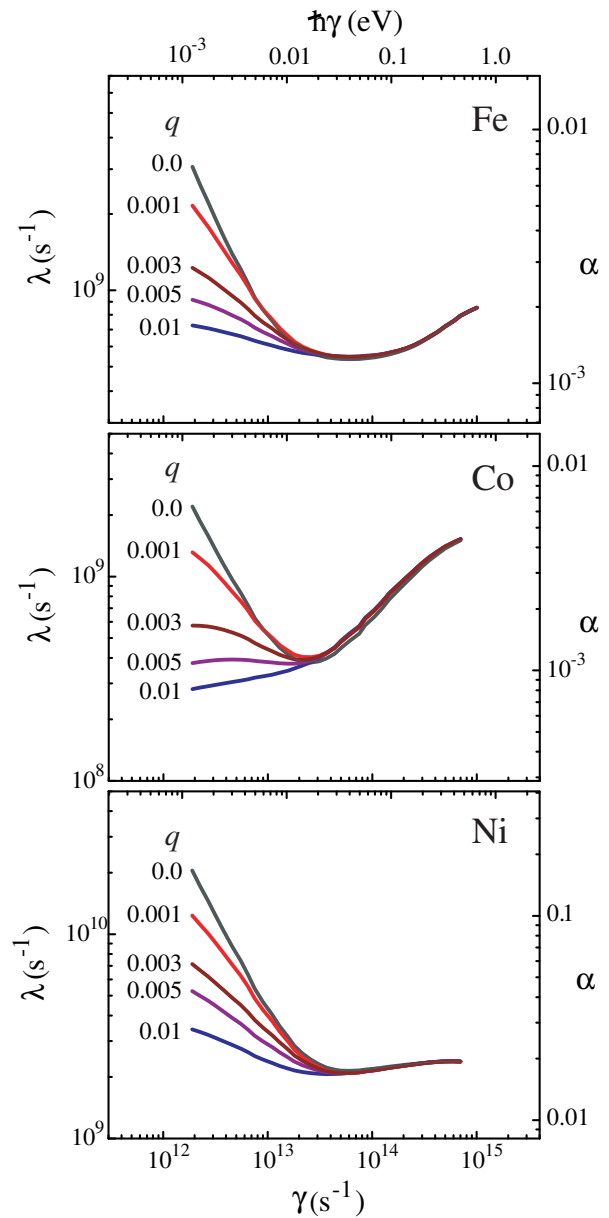}
 \caption{Damping rates versus scattering rate.  The precession damping rates for magnons
          in iron, cobalt, and nickel are plotted versus electron
          scattering rate for several magnon wavevectors.  A dramatic reduction in damping 
	  rate is observed at the lowest scattering rates.  The Landau-Lifshitz $\lambda$ 
          (Gilbert $\alpha$) damping parameter is given on the left (right) axes.  Electron 
          scattering rate is given in eV on the top axis.  Magnon wavevector magnitudes are 
          given in units of the Brillouin zone edge and directions are as indicated in the text.}
 \label{damping.total}
\end{figure}

Equation \ref{Kambersky.original}, which has been discussed
extensively \cite{Kambersky:1976, Gilmore:2007, Kambersky:2007, Gilmore:2008}, considers only local 
contributions to the damping rate.  Non-local contributions to damping may be studied through the decay of 
non-uniform spin-waves.  Although recent efforts have approached the problem of 
non-local contributions to the dissipation of 
non-collinear excited states \cite{Fahnle:2006, Brataas:2008} the simple step of generalizing 
Kambersk\'{y}'s theory to non-uniform mode magnons has not yet been taken.  We fill this obvious gap, 
obtaining a damping rate of
~
\begin{equation}
\lambda_{\bf q} = \frac{\pi \hbar \gamma_{M}^2}{\mu_0} \sum_{nm} \int d{\bf k} \left |
\Gamma^-_{nm}({\bf k}, {\bf k}+{\bf q}) \right |^2 W_{nm}({\bf k},{\bf k}+{\bf q}) \,
\label{Kambersky.generalized}
\end{equation}

\noindent for a magnon with wavevector ${\bf q}$.  We test the importance of non-local 
effects by quantifying this expression for varying degrees of magnetic 
non-collinearity (magnon wavevector magnitude).  The numerical evaluation of 
Eq.~\ref{Kambersky.generalized} for the damping rate of finite wavelength magnons 
in transition metal systems, presented in Fig.~\ref{damping.total}, and the ensuing physical discussion 
form the primary contribution of this paper.  We find that the damping rate expected in very pure 
samples at low temperature is rapidly reduced as the magnon wavevector $|{\bf q}|$ grows, but the 
damping rate anticipated outside of this ideal limit is barely affected.  We provide a simple band 
structure argument to explain these observations.  The results are relevant to systems for which the 
non-collinear excitation may be expanded in long wavelength spin-waves, provided the amplitude of 
these waves is small enough to neglect magnon-magnon scattering.


Calculations for the single-mode damping constant (Eq.~\ref{Kambersky.generalized}) as a 
function of electron scattering rate are presented in Fig.~\ref{damping.total} for iron, cobalt, and 
nickel.  The Gilbert damping parameter $\alpha = \lambda/\gamma_{\rm M}$ is also given.  Damping rates 
are given for magnons with wavevectors along the bulk equilibrium directions, 
which are $\langle 100 \rangle$ for Fe, $\langle 0001 \rangle$ for Co, and $\langle 111 \rangle$ for Ni.  
Qualitatively and quantitatively similar results were obtained for other magnon wavevector
directions for each metal.  The magnons reported on in Fig.~\ref{damping.total} constitute small 
deviations of the magnetization transverse to the equilibrium direction with wavevector magnitudes 
between zero and 1 \% of the Brillouin zone edge.  This wavevector range corresponds to magnon
half-wavelengths between infinity and 100 lattice spacings, which is 28.7 nm for Fe, 40.7 nm for Co, and 
35.2 nm for Ni.  This range includes the wavelengths reported by Vlaminck and Bailleul in their 
recent measurement of spin polarization \cite{Vlaminck:2008}.

Results for the three metals are qualitatively similar.  The most striking trend is a 
dramatic, order of magnitude decrease of the damping rate at the lowest scattering rate
tested as the wavevector magnitude increases from zero to 1 \% of the Brillouin zone edge.  This 
observation holds in each metal for every magnon propagation direction investigated.
For the higher scattering rates expected in devices at room temperature there is 
almost no change in the damping rate as the magnon wavevector increases from zero to 1 \% of the 
Brillouin zone edge in any of the directions investigated for any of the metals.

To understand the different dependences of the damping rate on the magnon wavevector at low 
versus high scattering rates we first note that the 
damping rate (Eqs.~\ref{Kambersky.original} \& \ref{Kambersky.generalized}) is 
a convolution of two factors: the torque matrix elements and the weighting 
function.  The matrix elements do not change significantly as the magnon wavevector 
increases, however, the weighting function can change substantially.  The weighting 
function
~
\begin{equation}
W_{nm}({\bf k},{\bf k}+{\bf q}) \approx A_{n,{\bf k}}(\epsilon_{\rm F}) A_{m,{\bf k}+{\bf 
q}}(\epsilon_{\rm F})
\end{equation}

\noindent contains a product of the initial and final state electron spectral 
functions
~
\begin{equation}
A_{n,{\bf k}}(\epsilon) = \frac{1}{\pi} \frac{\hbar\gamma}{(\epsilon - 
\epsilon_{n,{\bf k}})^2 + (\hbar\gamma)^2} \, ,
\end{equation}

\noindent which are Lorentzians in energy space.  The spectral function for state $\ket{n,{\bf k}}$, 
which has nominal band energy $\epsilon_{n,{\bf k}}$, is evaluated within a very narrow range of the 
Fermi level $\epsilon_{F}$.  The width of the spectral function $\hbar\gamma$ is given by the electron 
scattering rate $\gamma = 1/2\tau$ where $\tau$ is the orbital lifetime.  (The lifetimes of all orbital 
states are taken to be equal for these calculations and no specific scattering mechanism is implied.)  
The weighting function restricts the electron-hole pair 
generated during the magnon decay to states close in energy to each other and near 
the Fermi level.  For high scattering rates, the electron spectral functions are significantly 
broadened and the weighting function incorporates states within an appreciable 
range (several hundred meV) of the Fermi level.  For low scattering rates, the spectral functions are 
quite narrow (only a few meV) and both the electron and hole state must be very close to the Fermi 
level.  

\begin{figure}
 \includegraphics[angle=0,width=8.0cm]{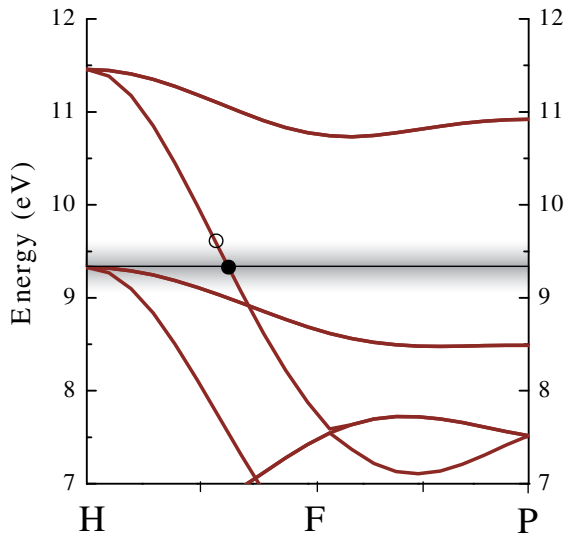}
 \caption{Partial band structure of bcc iron.  The horizontal black line 
          indicates the Fermi level and the shaded region represents the degree of spectral 
          broadening.  The solid dot is a hypothetical initial electron state while the open 
          circle is a potential final scattering state.  (Initial and final state wave-vector 
          separations are exaggerated for clarity of illustration.)  The intraband magnon decay 
          rate diminishes as the energy separation of the states exceeds the spectral broadening.}
 \label{band.structure}
\end{figure}

The second consideration useful for understanding the results of 
Fig.~\ref{damping.total} is that the sum in Eqs.~\ref{Kambersky.original} \& 
\ref{Kambersky.generalized} can be divided into intraband ($n=m$) and interband ($n \neq m$) terms.  
For the uniform mode, these two contributions correspond to different physical processes with the intraband contribution 
dominating at low scattering rates and the interband terms dominating at high scattering rates 
\cite{Kambersky:1976, Gilmore:2007, Kambersky:2007, Gilmore:2008}.  

\begin{figure}
 \includegraphics[angle=0,width=8.0cm]{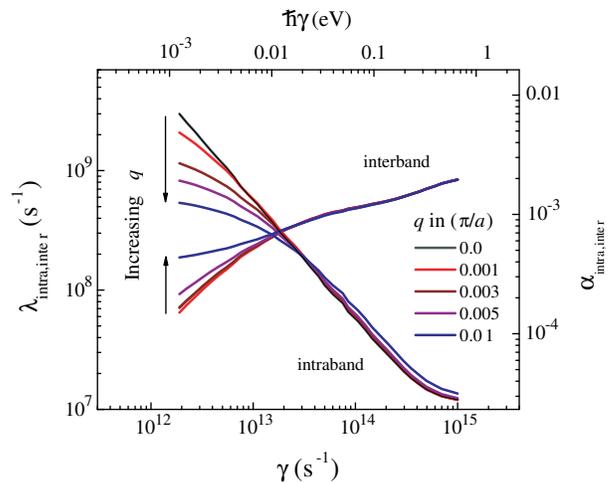}
 \caption{Intraband and interband damping contributions in iron.  The intraband and interband 
          contributions to the damping rate of magnons in the $\langle 100 \rangle$ direction in iron 
          are plotted versus scattering rate for several magnitudes of magnon wavevector.  Magnitudes 
          are given in units of the Brillouin zone edge.}
 \label{intra.inter}
\end{figure}

For intraband scattering, the electron and hole occupy the same band and must have 
essentially the same energy (within $\hbar \gamma$).  The energy difference between the electron and 
hole states may be approximated as $\epsilon_{n,{\bf k}+{\bf q}} - \epsilon_{n,{\bf k}} 
\approx {\bf q} \cdot \partial \epsilon_{n,{\bf k}} / \partial {\bf k}$.  The generation of intraband
electron-hole pairs responsible for intraband damping gets suppressed as ${\bf q} \cdot \partial 
\epsilon_{n,{\bf k}} / \partial {\bf k}$ becomes large compared to $\hbar \gamma$.
Unless the bands are very flat at the Fermi level there will be few locations on the Fermi surface that 
maintain the condition ${\bf q} \cdot \partial \epsilon_{n,{\bf k}} / \partial {\bf k} < \hbar \gamma$
for low scattering rates as the magnon wavevector grows.  (See Fig.~\ref{band.structure}).  Indeed, 
at low scattering rates when $\hbar \gamma$ is only a few meV, 
Fig.~\ref{intra.inter} shows that the intraband contribution to damping decreases markedly with only 
modest increase of the magnon wavevector.  Since the intraband contribution dominates the interband term 
in this limit the total damping rate also decreases sharply as the magnon wavevector is increased for 
low scattering rates.  For higher scattering rates, the electron spectral functions are
sufficiently broadened that the overlap of intraband states does not decrease
appreciably as the states are separated by finite wavevector (${\bf q} \cdot \partial \epsilon_{n,{\bf 
k}} / \partial {\bf k} < \hbar \gamma$ generally holds over the Fermi surface).  Therefore, the
intraband contribution is largely independent of magnon wavevector at high scattering rates.

The interband contribution to damping involves scattering between states in different bands, 
separated by the magnon wavevector ${\bf q}$.  Isolating the interband damping contribution 
reveals that these contributions are insensitive to the magnon wavevector at higher scattering 
rates where they form the dominant contribution to damping (see Fig.~\ref{intra.inter}).  To understand 
these observations we again compare the spectral broadening $\hbar \gamma$ to the quasiparticle energy 
difference $\Delta_{n,{\bf k}}^{m,{\bf k}+{\bf q}} = \epsilon_{m,{\bf k}+{\bf q}} - \epsilon_{n,{\bf 
k}}$.  The quasiparticle energy difference may be approximated as $\Delta_{n,{\bf k}}^{m,{\bf k}} + {\bf 
q} \cdot \partial \Delta_{n,{\bf k}}^{m,{\bf k}} / \partial {\bf k}$.  The interband energy spacings are 
effectively modulated by the product of the magnon 
wavevector and the slopes of the bands.  At high scattering rates, when the spectral broadening 
exceeds the vertical band spacings, this energy modulation is unimportant and the damping rate is 
independent of the magnon wavevector.  At low scattering rates, when the spectral broadening is less 
than many of the band spacings, this modulation can alter the interband energy spacings enough to 
allow or forbid generation of these electron-hole pairs.  For Fe, Co, and Ni, this produces a 
modest increase in the interband damping rate at low scattering rates as the magnon wavevector 
increases.  However, this effect is unimportant to the total damping rate, which remains dominated 
by the intraband terms at low scattering rates.


Lastly, we describe the numerical methods employed in this study.  Converged ground state electron 
densities were first obtained via the linear-augmented-plane-wave method.  The Perdew-Wang functional 
for exchange-correlation within the local spin density approximation was implemented.  Many details 
of the ground state density convergence process are given in \cite{Mattheiss.Hamann:1986}.
Densities were then expanded into Kohn-Sham orbitals using a scalar-relativistic spin-orbit
interaction with the magnetization aligned along the experimentally determined magnetocrystalline
anisotropy easy axis.  The Kohn-Sham energies were artificially broadened through the {\em ad hoc}
introduction of an electron lifetime.  Matrix elements of the torque operator $\Gamma^{-} = [\sigma^-,
{\cal H}_{\rm so}]$ were evaluated similarly to the spin-orbit matrix elements \cite{Stiles:2001}.
($\sigma^-$ is the spin lowering operator and ${\cal H}_{\rm so}$ is the spin-orbit Hamiltonian.)  The
product of the matrix elements and the weighting function were integrated over $k$-space using the
special points method with a fictitious smearing of the Fermi surface for numerical stability.
Convergence was obtained by sampling the full Brillouin zone with $160^3$ $k$-points for Fe and 
Ni, and $160^2$ x $91$ points for Co.


In summary, we have investigated the importance of non-local damping effects by calculating
the intrinsic spin-orbit contribution to precession damping in bulk 
transition metal ferromagnets for small amplitude spin-waves with finite wavelengths.  
Results of the 
calculations do not contradict the common-practice assumption that damping is a local phenomenon.  For 
transition metals, at 
scattering rates corresponding to room temperature, we find that the single-mode damping rate is 
essentially independent of magnon wavevector for wavevectors between zero and 1 \% of the Brillouin zone 
edge.  It is not until low temperatures in the most pure 
samples that non-local effects become significant.  At these scattering rates, damping rates decrease 
by as much as an order of magnitude as the magnon wavevector is increased.  The insensitivity of 
damping rate to magnon wavevector at high scattering rates versus the strong sensitivity at low 
scattering rates can be explained in terms of band structure effects.  Due to electron spectral 
broadening at high scattering rates the energy conservation constraint during magnon 
decay is effectively relaxed, making the damping rate independent of magnon wavevector.  The minimal 
spectral broadening at low scattering rates -- seen only in very pure and cold samples -- greatly 
restricts the possible intraband scattering processes, lowering the damping rate.
The prediction of reduced damping at low scattering rates and non-zero magnon 
wavevectors is of little practical importance, but could provide an accessible test of the 
torque-correlation model.  Specifically, this might be testable in ferromagnetic semiconductors such as 
(Ga,Mn)As for which many spin-wave resonances have been experimentally observed at low temperatures 
\cite{Goennenwein:2003}.

This work has been supported in part through NIST-CNST / UMD-Nanocenter cooperative agreement.

\end{document}